\input harvmac.tex
%
%
\noblackbox


\def\inbar{\,\vrule height1.5ex width.4pt depth0pt}
\def\IB{\relax{\rm I\kern-.18em B}}
\def\IC{\relax\hbox{$\inbar\kern-.3em{\rm C}$}}
\def\ID{\relax{\rm I\kern-.18em D}}
\def\IE{\relax{\rm I\kern-.18em E}}
\def\IF{\relax{\rm I\kern-.18em F}}
\def\IG{\relax\hbox{$\inbar\kern-.3em{\rm G}$}}
\def\IH{\relax{\rm I\kern-.18em H}}
\def\II{\relax{\rm I\kern-.18em I}}
\def\IK{\relax{\rm I\kern-.18em K}}
\def\IL{\relax{\rm I\kern-.18em L}}
\def\IM{\relax{\rm I\kern-.18em M}}
\def\IN{\relax{\rm I\kern-.18em N}}
\def\IO{\relax\hbox{$\inbar\kern-.3em{\rm O}$}}
\def\IP{\relax{\rm I\kern-.18em P}}
\def\IQ{\relax\hbox{$\inbar\kern-.3em{\rm Q}$}}
\def\IR{\relax{\rm I\kern-.18em R}}
\font\cmss=cmss10 \font\cmsss=cmss10 at 7pt
\def\IZ{\relax\ifmmode\mathchoice
{\hbox{\cmss Z\kern-.4em Z}}{\hbox{\cmss Z\kern-.4em Z}}
{\lower.9pt\hbox{\cmsss Z\kern-.4em Z}}
{\lower1.2pt\hbox{\cmsss Z\kern-.4em Z}}\else{\cmss Z\kern-.4em
Z}\fi}
\def\IGa{\relax\hbox{${\rm I}\kern-.18em\Gamma$}}
\def\IPi{\relax\hbox{${\rm I}\kern-.18em\Pi$}}
\def\ITh{\relax\hbox{$\inbar\kern-.3em\Theta$}}
\def\IOm{\relax\hbox{$\inbar\kern-3.00pt\Omega$}}

\font\zfont = cmss10 
\font\litfont = cmr6

\def\bigone{\hbox{1\kern -.23em {\rm l}}}
\def\ZZ{\hbox{\zfont Z\kern-.4emZ}}

\def\half{{\litfont {1 \over 2}}}

\def\IR{\relax{\rm I\kern-.18em R}}
\def\Dsl{\,\raise.15ex\hbox{/}\mkern-13.5mu D} 
\def\Gsl{\,\raise.15ex\hbox{/}\mkern-13.5mu G} 
\def\Csl{\,\raise.15ex\hbox{/}\mkern-13.5mu C} 
\font\cmss=cmss10 \font\cmsss=cmss10 at 7pt
%
%
\lref\ppvn{M. Pernici, K. Pilch and P. van Nieuwenhuizen, ``Gauged Maximally
Extended Supergravity in Seven Dimensions,''  Phys. Lett.
{\bf B143} (1984) 103.}
\lref\ppvnw{M. Pernici, K. Pilch,  P. van Nieuwenhuizen and N. P. Warner,
``Noncompact Gaugings and Critical Points of Maximal Supergravity in
Seven-dimensions,''  Nucl. Phys. {\bf B249} (1985) 381.}
\lref\largen{J. Maldacena, ``The Large N Limit of Superconformal Field
Theories and Supergravity,'' Adv. Theor. Math. Phys. {\bf 2} (1998) 231.}
\lref\wittbar{E. Witten, ``Baryons And Branes In Anti de Sitter Space,''
JHEP {\bf 07} (1998) 006.}
\lref\wtpt{E. Witten, ``Anti-de Sitter Space, Thermal Phase Transition, and
Confinement in Gauge Theories," Adv. Theor. Math. Phys. {\bf 2}
(1998) 505.}
\lref\groo{D. J. Gross and H. Ooguri, ``Aspects of Large N Gauge Theory
Dynamics as Seen by String Theory,''
Phys. Rev. {\bf D58} (1998) 106002.}
\lref\lpt{H. Lu, C. N. Pope and P. K. Townsend, ``Domain Walls from Anti-de
Sitter Spacetime,''  Phys. Lett. {\bf B391} (1997) 39.}
\lref\wittfive{E. Witten, ``Five-Brane Effective Action In M-Theory,''
hep-th/9610234.}
\lref\ahaw{O. Aharony and E. Witten, ``Anti-de Sitter Space
and the Center of the Gauge Group,''
hep-th/9807205.}
\lref\afgj{D. Anselmi, D.Z. Freedman, M.T. Grisaru, and A.A. Johansen,
``Nonperturbative formulas for central functions of supersymmetric
gauge theories,''   hep-th/9708042}
\lref\dlm{M.J. Duff, J.T. Liu and R. Minasian,``Eleven dimensional
origin of string/string duality: a one loop test,'' Nucl. Phys. {\bf B452}
(1995) 261, hep-th/9506126.}
\lref\wittads{E. Witten, ``Anti de Sitter space and holography,''
hep-th/9802150.}
\lref\hst{P. Howe, G. Sierra, and P. Townsend,
``Supersymmetry in six dimensions,''  Nucl. Phys.
{\bf B221} (1983) 331.}
\lref\krvn{M. Gunaydin, L. J. Romans and N. P. Warner, ``Gauged $N=8$
supergravity in five dimensions,''  Phys. Lett. {\bf B154} (1985) 268\semi H.J.
Kim ,  L.J. Romans, P. van Nieuwenhuizen,
``The mass spectrum of chiral $N=2,D=10$ supergravity
on $S^5$,'' Phys. Rev. {\bf D32} (1985) 389\semi M. Pernici, K. Pilch and P.
van Nieuwenhuizen,
``Gauged $N=8$, $d=5$ supergravity,''  Nucl. Phys. {\bf B259} (1985) 460. }
\lref\fhmm{D. Freed, J.A. Harvey, R. Minasian and G. Moore, ``Gravitational
Anomaly Cancellation for  $M$-Theory Fivebranes,''  hep-th/9803205.}
\lref\bottcatt{R. Bott and A.S. Cattaneo, ``Integral
invariants of 3-manifolds,'' dg-ga/9710001.}
\lref\clash{A. Losev, G. Moore, N. Nekrasov, and
S. Shatashvili, ``Chiral Lagrangians, Anomalies, Supersymmetry,
and holomorphy,'' Nucl. Phys. {\bf B484} (1997)196, hep-th/9606082.}
\lref\klebs{ I.R. Klebanov and A.A. Tseytlin, `` Entropy of Near-Extremal Black
$p$-branes,'' Nucl. Phys. {\bf B475} (1996) 164, hep-th/9604089\semi  S.S.
Gubser and I.R. Klebanov, ``Absorption by Branes
and Schwinger Terms in the World Volume Theory,'' Phys. Lett.
{\bf B413} (1997) 41, hep-th/9708005.}
\lref\hennsken{M. Henningson and K. Skenderis, ``The Holographic Weyl
Anomaly,'' hep-th/9806087.}
\lref\msw{J. Maldacena, A. Strominger and E. Witten, ``Black Hole Entropy
in M Theory,'' J. High Energy Phys. 12 (1997) 2, hep-th/9711053.}
\lref\freund{P.G.O. Freund and M. A. Rubin, ``Dynamics of Dimensional
Reduction,'' Phys. Lett. {\bf 97B} (1980) 233.}
\lref\kkrefs{
M.J. Duff, B.E.W. Nilsson and C.N. Pope, ``Kaluza-Klein Supergravity,''
Phys. Rep. 130 (1986) 1.}
\lref\gkp{S. Gubser, I. Klebanov and A. Polyakov,
``Gauge Theory Correlators
from Noncritical String Theory,'' Phys. Lett. {\bf B428}
(1998) 105, hep-th/9802109.}
\lref\GrHa{P.~ Griffiths and J.~ Harris, {\it Principles of
Algebraic Geometry},  J.Wiley and Sons, 1978. }
\lref\vwrsq{C. Vafa and E. Witten, ``A One-loop Test of String Duality,''
Nucl. Phys. {\bf B447} (1995) 261, hep-th/9505053.}
\lref\cnss{D.Z. Freedman, S.D. Mathur, A. Matusis and L. Rastelli,,
``Correlation functions in the $CFT(d)/AdS(d+1)$ correspondence,''
hep-th/9804058\semi G. Chalmers, H. Nastase, K. Schalm and R. Siebelink,
``R-Current Correlators in $N=4$ Super Yang-Mills Theory from
Anti-de Sitter Supergravity,'' hep-th/9805105.}
\lref\pvnt{K. Pilch, P. van Nieuwenhuizen and P.K. Townsend,
``Compactification of $D=11$ supergravity on
$S(4)$ (or $11=7+4$, too),''  Nucl. Phys. {\bf B242} (1984) 377.}
\lref\acf{{\it Modern
Kaluza-Klein Theories},
T. Appelquist, A. Chodos, and P.G.O. Freund, eds.
Frontiers in Physics v. 65, Addison-Wesley, 1987.}
\lref\greenco{M.B. Green and P. Vanhove, ``D-instantons, Strings and
M-theory,''  Phys.Lett. {\bf B408} (1997) 122, hep-th/9704145\semi M.B. Green,
M. Gutperle and P. Vanhove, ``One loop in eleven dimensions,''  Phys.Lett. {\bf
B409} (1997) 177, hep-th/9706175.}
\lref\afmn{I. Antoniadis, S. Ferrara, R.~Minasian and K.S. Narain, ``$R^4$
Couplings in M and Type II Theories on Calabi-Yau Spaces,'' Nucl. Phys. {\bf B
507} (1997) 571; hep-th/9707013.}
\lref\tpvn{P.K.~Townsend, K.~Pilch and P.~van Nieuwenhuizen,
``Selfduality In Odd Dimensions," Phys. Lett. {\bf 136B} (1984) 38.}
\lref\adsbh{M. Duff and J.T. Liu, ``Anti-de Sitter Black Holes in Gauged
$N=8$ Supergravity,'' hep-th/9901149.}
\lref\romans{L.J. Romans, ``Supersymmetric, cold and lukewarm black holes in
cosmological Einstein-Maxwell theory," Nucl. Phys. {\bf B383} (1992) 395.}
\lref\bcs{K. Behrndt, A.H. Chamseddine and W.A. Sabra, ``BPS black holes in
$N=2$ five-dimensional AdS supergravity," Phys. Lett. {\bf B442} (1998)
97.}
\lref\bcvs{K. Behrndt, M. Cveti\v{c} and W.A. Sabra, ``Non-extreme black
holes of five-dimensional $N=2$ AdS supergravity," hep-th/9810227.}
\lref\cdk{M.M. Caldarelli and D. Klemm, ``Supersymmetry of Anti-de Sitter
Black Holes," hep-th/9808097.}
\lref\dk{D. Klemm, ``BPS Black Holes in Gauged $N=4$, $D=4$ Supergravity,"
hep-th/9810090.}
\lref\was{W.A. Sabra, ``Anti-De Sitter BPS Black Holes in $N=2$ Gauged
Supergravity," hep-th/9903143.}
\lref\cg{M. Cveti\v{c} and S.S. Gubser, ``Phases of R-charged Black Holes,
Spinning Branes and Strongly Coupled Gauge Theories," hep-th/9902195.}
\lref\lpr{H. L\"u, C.N. Pope and J. Rahmfeld, ``A Construction of Killing
Spinors on $S^n$," hep-th/9805151.}
\lref\ferp{S. Ferrara and M. Porrati, ``$AdS_5$ Superalgebras with Brane
Charges," hep-th/9903241.}
\lref\tena{M. Cveti\v{c} et al., ``Embedding AdS Black Holes in Ten and
Eleven Dimensions," hep-th/9903214.}
\lref\intbr{A. Tseytlin, ``Harmonic superpositions of M-branes,"
Nucl. Phys. {\bf B475} (1996) 149.}
\lref\jpg{J.P. Gauntlett, ``Intersecting branes," in {\it Seoul/Sokcho 1997,
Dualities in gauge and string theories}, hep-th/9705011.}
\lref\youm{D. Youm, ``Localized intersecting BPS branes," hep-th/9902208.}
\lref\ity{N.~Itzhaki, A.A.~Tseytlin and S.~Yankielowicz,
``Supergravity solutions for branes localized within branes,"
Phys. Lett. {\bf B432} (1998) 298.}
%
%
%

\Title{\vbox{\baselineskip12pt
\hbox{RU99-5-B}
\hbox{YCTP-P9-99}
\hbox{hep-th/9903269}
}}
{\vbox{\centerline{
Black holes and membranes in $AdS_7$}
}}

\centerline{James T. Liu}
\medskip
\centerline{Department of Physics, The Rockefeller University}
\centerline{1230 York Avenue, New York, NY 10021--6399}

\bigskip
\centerline{and}
\bigskip

\centerline{Ruben Minasian}
\medskip
\centerline{Department of Physics, Yale University}
\centerline{New Haven, CT 06520}

\bigskip
\centerline{\bf Abstract}

We investigate maximal gauged supergravity in seven dimensions and some of
its solitonic solutions.  By focusing on a truncation of the gauged
$SO(5)$ $R$-symmetry group to its $U(1)^2$ Cartan subgroup, we construct
general two charge black holes that are asymptotically anti-de Sitter.  We
demonstrate that 1- and 2-charge black holes preserve ${1\over2}$ and
${1\over4}$ of the supersymmetries respectively.  Additionally, we examine
the odd-dimensional self-duality equation governing the three-form potential
transforming as the $5$ of $SO(5)$, and provide some insight on the
construction of membrane solutions in anti-de Sitter backgrounds.

\Date{March 1999}

%
\newsec{Introduction}
The gauged supergravities in odd dimensions have an interesting feature:
the anti-symmetric tensor potentials transforming in the fundamental
representation of the $R$-symmetry group obey first order self-duality
conditions \refs{\tpvn,\ppvn,\pvnt}.  In particular, this odd-dimensional
self-duality construction was shown to be necessary in order to ensure that
only the correct propagating modes of the anti-symmetric tensor potentials
survives in the presence of gauging.  In this note we focus on the gauged
$N=4$ ({\it i.e.\ }the maximally supersymmetric) theory in seven dimensions
\ppvn, although similar results may be extended to other odd-dimensional
theories as well.

The $N=4$ gauged seven-dimensional supergravity theory involves the gauging
of a $SO(5)$ subgroup of $SL(5,R)$, resulting in a theory with both a gauged
$SO(5)_g$ and a composite $SO(5)_c$ local invariance.  The bosonic field
content includes a graviton, ten Yang-Mills gauge fields transforming in the
adjoint of $SO(5)_g$, five rank-three tensors transforming as a $5$
under $SO(5)_g$ and 14 scalars parameterizing a $SL(5, R)/SO(5)_c$ coset.
For the fermions, there are four gravitini and 16 spin-${1\over2}$ fields
transforming as the $4$ and $16$ of $SO(5)_c$ respectively.
Interestingly enough, the potential arising from the gauging has
two critical points---a saddle point with $SO(4)$ symmetry and a $SO(5)$
stationary point.  Without brany ``matter'', the former is unstable and
breaks all supersymmetries \ppvnw.

Our goal is to analyze solitonic objects charged under the
$SO(5)_g$ $R$-symmetry group.  These include both 0- and 3-branes
carrying vector-potential charges and 1- and 2-branes charged with
respect to the three-form potentials. The latter objects have been
anticipated (see {\it e.g.\ }\refs{\wittbar, \groo}) and their
existence plays an important role in the different aspects of
$AdS$/CFT correspondence \largen. For a recent discussion of branes
in $AdS$ spaces from the point of view of BPS algebras in the
$AdS_5$ case see \ferp. In particular, Ref.~\ferp\ examines the
conditions imposed by the preservation of half or less supersymmetry on
the $R$-symmetry group in $AdS_5$.  Although similar analysis is lacking
for $AdS_7$, we show in this note that (at least at the level
of solutions) $SO(5) \rightarrow SO(4)$ is required in $AdS_7$
for having a membrane which is electrically charged with respect to
the three-form potential and with gauge fields forming an instanton
configuration transverse to the brane. It would be interesting to find the
corresponding extension of the supersymmetry algebra and the complete
analysis of the branes in this case.

\newsec{$N=4$, $D=8$ gauged supergravity}
We begin with a discussion of the gauged $N=4$, $D=7$ supergravity theory
itself.  The bosonic Lagrangian takes the form \ppvn:
\eqn\action{\eqalign{2\kappa^2e^{-1}{\cal L} = 
R& + \half m^2 (T^2 - 2T_{ij}T^{ij}) - Tr(P_{\mu} P^{\mu})
- \half (V_I{}^i V_J{}^j F_{\mu \nu}^{IJ})^2
+ m^2 (V_i^{-1\,I} C_{\mu \nu \rho}^I)^2 \cr
&+ e^{-1}\Bigl( {1 \over 2} \delta^{IJ} (C_3)_ I \wedge (dC_3)_J
+ m \epsilon_{IJKLM} (C_3)_I F_2^{JK} F_2^{LM}
+ m^{-1} p_2(A, F) \Bigr). \cr}}
Here $I,J = 1, \ldots, 5$ denote $SO(5)_g$ indices, and $i,
j = 1, \ldots, 5$ denote $SO(5)_c$ indices.  In general we follow the
notation of Ref.~\ppvn, except that we work with a metric of signature
$(-+\ldots+)$.  The 14 scalar degrees of freedom
are contained in the $SL(5, R)/SO(5)$ coset element $V_I{}^i$, transforming as
a $5$ under both $SO(5)_g$ and $SO(5)_c$.  The scalar kinetic term,
$P_\mu^{ij}$, and composite $SO(5)_c$ connection, $Q_\mu{}^i{}_j$, are
defined through $ V_i^{-1\,I} {\cal D}_{\mu}
V_I{}^j = \left( Q_{\mu} \right)_{[ij]} + \left( P_{\mu}
\right)_{(ij)}$, where ${\cal D}_\mu$ is a fully gauge covariant derivative
so that {\it e.g.\ }${\cal D}_\mu V_I{}^j=\partial_\mu V_I{}^j+A_{\mu\, I}^J
V_J{}^j$.  Finally, the $T$-tensor is defined by $T_{ij} = V_i^{-1\,I}
V_j^{-1\,J} \delta_{IJ}$, and $T=T_{ij} \delta^{ij}$.

The fermionic supersymmetries associated with the lagrangian \action\ take on
the form
\eqn\grsusy{\eqalign{
\delta\psi_\mu=\Bigl[{\cal D}_\mu+{m\over20}T\gamma_\mu
&-{1\over40}(\gamma_\mu{}^{\nu\lambda}-8\delta_\mu^\nu\gamma^\lambda)
\Gamma^{ij}V_I{}^iV_J{}^jF_{\nu\lambda}^{IJ}\cr
&+{m\over10\sqrt{3}}(\gamma_\mu{}^{\nu\lambda\sigma}
-{\textstyle{9\over2}}\delta_\mu^\nu
\gamma^{\lambda\sigma})\Gamma^iV_i^{-1\,I}C_{\nu\lambda\sigma}^I
\Bigr]\epsilon,
}}
for the gravitini, and
\eqn\disusy{\eqalign{
\delta\lambda_i=\Bigl[{m\over2}(T_{ij}-{\textstyle{1\over5}}
\delta_{ij}T)\Gamma^j
+{1\over2}\gamma^\mu P_{\mu\,ij}\Gamma^j
&+{1\over16}\gamma^{\mu\nu}(\Gamma^{kl}\Gamma^i-{\textstyle{1\over5}}\Gamma^i
\Gamma^{kl})V_K{}^kV_L{}^lF_{\mu\nu}^{KL}\cr
&+{m\over20\sqrt{3}}\gamma^{\mu\nu\lambda}(\Gamma^{ij}-4\delta^{ij})
V_j^{-1\,J}C_{\mu\nu\lambda}^J\Bigr]\epsilon,
}}
for the spin-${1\over2}$ fermions $\lambda_i$.  Note that $\epsilon$
transforms as the $4$ (spinor) of $SO(5)_c$, with corresponding Dirac
matrices $\Gamma^i$.  Although such spinor indices are hidden for simplicity,
we note here that not only is $\lambda_i$ $\Gamma$-traceless,
$\Gamma^i\lambda_i=0$, but its variation \disusy\ is as well (as is required
for consistency).

Since the full gauged supergravity theory is rather involved, we proceed
with a simplification of the field content.  In particular, as $SO(5)_g$ has
rank two, we focus on a $U(1)^2$ truncation of the full non-abelian theory
(a similar simplification was performed in \adsbh\ in the context of the
$N=8$, $D=4$ gauged supergravity theory).  Performing a gauge choice of
identifying $SO(5)_g$ with $SO(5)_c$, we specialize to a diagonal scalar
vielbein of the form
\eqn\viel{V_I{}^i={\rm diag}[\matrix{e^{-\lambda_1}&e^{-\lambda_1}&
e^{-\lambda_2}&e^{-\lambda_2}&e^{2\lambda_1+2\lambda_2}}],}
where $\lambda_1$ and $\lambda_2$ are two independent real scalars.  With
this choice it is natural to restrict ourselves to the two Cartan gauge
fields $A_\mu^{12}\equiv A_\mu^{(1)}$ and $A_\mu^{34}\equiv A_\mu^{(2)}$ as
well as a single three-form potential $C_{\mu\nu\lambda}^5$ (we subsequently
drop the index 5 as there will be no confusion with only a single three-form
field).

With the above choice of truncation, we are left with a model containing
gravity, two scalars, two $U(1)$ gauge fields and a single three-form
potential.  While this field content does not necessarily correspond to a
{\it consistent} truncation of the $N=4$ model, solutions to the truncated
equations of motion will necessarily extend to solutions of the full theory.
With this in mind, the truncated bosonic lagrangian has the form
\eqn\trunc{\eqalign{
2\kappa^2e^{-1}{\cal L}=&R-{1\over2}m^2{\cal V}-5\partial_\mu(\lambda_1
+\lambda_2)^2-\partial_\mu(\lambda_1-\lambda_2)^2
-e^{-4\lambda_1}F_{\mu\nu}^{(1)\,2} -e^{-4\lambda_2}F_{\mu\nu}^{(2)\,2}\cr
&+m^2e^{-4\lambda_1-4\lambda_2}C_{\mu\nu\lambda}^{2}
-{m\over6}\epsilon^{\mu\nu\lambda\alpha\beta\gamma\delta}
C_{\mu\nu\lambda}\partial_\alpha C_{\beta\gamma\delta}\cr
&+{1\over\sqrt{3}}\epsilon^{\mu\nu\lambda\alpha\beta\gamma\delta}
C_{\mu\nu\lambda}F_{\alpha\beta}^{(1)}F_{\gamma\delta}^{(2)}
+m^{-1}p_2(A,F),}}
where the scalar potential ${\cal V}$ is
\eqn\spot{{\cal V}=-8e^{2\lambda_1+2\lambda_2}
-4e^{-2\lambda_1-4\lambda_2} -4e^{-4\lambda_1-2\lambda_2}
+e^{-8\lambda_1-8\lambda_2}.}
The resulting scalar equations of motion are
\eqn\seom{\eqalign{
\nabla^2(3\lambda_1+2\lambda_2)&=-e^{-4\lambda_1}F^{(1)\,2}+m^2
e^{-4\lambda_1-4\lambda_2}C^2+{m^2\over8}{\partial{\cal V}\over
\partial\lambda_1},\cr
\nabla^2(2\lambda_1+3\lambda_2)&=-e^{-4\lambda_2}F^{(2)\,2}+m^2
e^{-4\lambda_1-4\lambda_2}C^2+{m^2\over8}{\partial{\cal V}\over
\partial\lambda_2},}}
while the gauge equations of motion are
\eqn\geom{\eqalign{
\nabla^\mu(e^{-4\lambda_1}F_{\mu\nu}^{(1)})&={1\over2\sqrt{3}}
\epsilon_{\mu\nu}{}^{\lambda\sigma\alpha\beta\gamma}\nabla^\mu
(F_{\lambda\sigma}^{(2)}C_{\alpha\beta\gamma}),\cr
\nabla^\mu(e^{-4\lambda_2}F_{\mu\nu}^{(2)})&={1\over2\sqrt{3}}
\epsilon_{\mu\nu}{}^{\lambda\sigma\alpha\beta\gamma}\nabla^\mu
(F_{\lambda\sigma}^{(1)}C_{\alpha\beta\gamma}).}}
Note in particular how these equations mix $F^{(1)}$ and $F^{(2)}$.  The
three-form $C_{\mu\nu\lambda}$ satisfies the odd-dimensional self duality
equation
\eqn\ceom{e^{-4\lambda_1-4\lambda_2}C_{\mu\nu\lambda}={1\over6m}
\epsilon_{\mu\nu\lambda}{}^{\alpha\beta\gamma\delta}\partial_\alpha
C_{\beta\gamma\delta}-{1\over2\sqrt{3}m^2}
\epsilon_{\mu\nu\lambda}{}^{\alpha\beta\gamma\delta}F_{\alpha\beta}^{(1)}
F_{\gamma\delta}^{(2)}.}
Finally, the Einstein equation may be written in Ricci form
\eqn\eins{\eqalign{R_{\mu\nu}=&{m^2\over10}g_{\mu\nu}{\cal V}
+5\partial_\mu(\lambda_1+\lambda_2) \partial_\nu(\lambda_1+\lambda_2)
+\partial_\mu(\lambda_1-\lambda_2) \partial_\nu(\lambda_1-\lambda_2)\cr
&+2e^{-4\lambda_1}(F^{(1)\,2}_{\mu\nu}
-{\textstyle{1\over10}}g_{\mu\nu}F^{(1)\,2})
+2e^{-4\lambda_2}(F^{(2)\,2}_{\mu\nu}
-{\textstyle{1\over10}}g_{\mu\nu}F^{(2)\,2})\cr
&-3m^2e^{-4\lambda_1-4\lambda_2}(C^2_{\mu\nu}-{\textstyle{2\over15}}C^2).}}

For the above bosonic $U(1)^2$ ansatz, we find that the fermionic
supersymmetries, \grsusy\ and \disusy, simplify when written in terms
of a shifted gravitino, $\hat\psi_\mu\equiv \psi_\mu+{1\over2}
\gamma_\mu\Gamma^5\lambda_5$, and the
appropriate linear combinations of $\lambda_i$, $i=1,\ldots,5$:
\eqn\linc{\lambda^{(2)}={3\over2}\Gamma^1\lambda_1+\Gamma^3\lambda_3,
\qquad \lambda^{(1)}=\Gamma^1\lambda_1+{3\over2}\Gamma^3\lambda_3.}
The resulting supersymmetry variations are given by
\eqn\gvar{\eqalign{\delta\hat\psi_\mu=\Bigl[\nabla_\mu+{g\over2}
(A_\mu^{(1)}&\Gamma^{12}+A_\mu^{(2)}\Gamma^{34})
+{m\over4}e^{-4\lambda_1-4\lambda_2}\gamma_\mu
+{1\over2}\gamma_\mu\gamma^\nu\partial_\nu(\lambda_1+\lambda_2)\cr
&+{1\over2}\gamma^\nu(e^{-2\lambda_1}F_{\mu\nu}^{(1)}\Gamma^{12}
+e^{-2\lambda_2}F_{\mu\nu}^{(2)}\Gamma^{34})\cr
&-{m\sqrt{3}\over4}\gamma^{\nu\lambda}e^{-2\lambda_1-2\lambda_2}
C_{\mu\nu\lambda}\Gamma^5\Bigr]\epsilon,}}
and
\eqn\dvar{\eqalign{\delta\lambda^{(1)}=\Bigl[{m\over4}(e^{2\lambda_1}
-e^{-4\lambda_1-4\lambda_2})-{1\over4}\gamma^\mu\partial_\mu(3\lambda_1
&+2\lambda_2)-{1\over8}\gamma^{\mu\nu}e^{-2\lambda_1}F_{\mu\nu}^{(1)}
\Gamma^{12}\cr
&+{m\over8\sqrt{3}}\gamma^{\mu\nu\lambda}
e^{-2\lambda_1-2\lambda_2}C_{\mu\nu\lambda}\Gamma^5\Bigr]\epsilon,\cr
\delta\lambda^{(2)}=\Bigl[{m\over4}(e^{2\lambda_2}
-e^{-4\lambda_1-4\lambda_2})-{1\over4}\gamma^\mu\partial_\mu(2\lambda_1
&+3\lambda_2)-{1\over8}\gamma^{\mu\nu}e^{-2\lambda_2}F_{\mu\nu}^{(2)}
\Gamma^{34}\cr
&+{m\over8\sqrt{3}}\gamma^{\mu\nu\lambda}
e^{-2\lambda_1-2\lambda_2}C_{\mu\nu\lambda}\Gamma^5\Bigr]\epsilon.\cr
}}

\newsec{The AdS black hole solution}
Black hole solutions in spaces that are asymptotically either de
Sitter or anti-de Sitter have been known for a long time.  More
recently, supersymmetry properties of the four-dimensional
Reissner-Nordstrom-$AdS$ black hole have been considered in
\romans\ in the context of $N=2$ gauged supergravity (which
contains a single graviphoton and no scalars). Subsequently, more
general multiple charged $AdS$ black holes have been constructed
and studied in five dimensions \refs{\bcs,\bcvs} as well as in four
dimensions \refs{\cdk\dk\adsbh\was}.

All such multiple-charged $AdS$ black holes have a similar
structure regardless of dimension, and appear as simple $AdS$
generalizations of the $p=0$ case of the family of generic
$p$-brane solutions. This in fact led the authors of \cg\ to
present the $AdS$ black hole solution of $N=4$, $D=7$ gauged
supergravity without actual construction of the lagrangian
\trunc.  Presently we construct a two-charge $AdS$ black hole solution to the
bosonic equations of motion \seom--\eins\ after setting the three-form
potential to zero as appropriate for a black hole.  The resulting general
non-extremal solution has the form
\eqn\twoch{\eqalign{
&ds^2=-(H_1H_2)^{-4/5}f\,dt^2+(H_1H_2)^{1/5}\Bigl({dr^2\over f}
+r^2d\Omega_5^2\Bigr),\cr
&A_0^{(\alpha)}={\eta_\alpha\over2}\coth\mu_\alpha (H_\alpha^{-1}-1),\cr
&e^{3\lambda_1+2\lambda_2}=H_1^{-1/2},\qquad
e^{2\lambda_1+3\lambda_2}=H_2^{-1/2},
}}
where $H_1$ and $H_2$ are harmonic functions,
\eqn\harm{H_\alpha=1+{k\sinh^2\mu_\alpha\over r^4},}
and $\eta_\alpha=\pm1$ sets the sign of the electric charge.  The function
$f$ is given by
\eqn\eff{f=1-{k\over r^4}+{m^2r^2\over4}H_1H_2,}
and contains information on both the non-extremality of the solution
and its anti de-Sitter nature.
The extremal limit is obtained by letting $k\to0$ and $\mu_\alpha\to\infty$
with $Q_\alpha\equiv k\sinh^2\mu_\alpha$ fixed.

Turning to the supersymmetry properties of the extremal black holes, we
find it more convenient to begin with the spin-${1\over2}$ variations \dvar.
In the extremal limit we find
\eqn\sdsusy{\eqalign{
\delta\lambda^{(1)}&={1\over8}{\cal H}^{-1/10}\partial_r\log H_1
\gamma^{\overline r}[f^{1/2}-\eta_1\gamma_{\overline{0}}\Gamma^{12}
+{mr\over2}{\cal H}^{1/2}\gamma_{\overline r}]\epsilon,\cr
\delta\lambda^{(2)}&={1\over8}{\cal H}^{-1/10}\partial_r\log H_2
\gamma^{\overline r}[f^{1/2}-\eta_2\gamma_{\overline{0}}\Gamma^{34}
+{mr\over2}{\cal H}^{1/2}\gamma_{\overline r}]\epsilon,}}
with ${\cal H}=H_1H_2$.  This indicates the presence of two individual
${1\over2}$-supersymmetry projections:
\eqn\sproj{P^{(\alpha)}_{\eta_\alpha}={1\over2}[1-f^{-1/2}(
-\eta_\alpha\gamma_{\overline 0}\Gamma^{(\alpha)}+{mr\over2}{\cal H}^{1/2}
\gamma_{\overline r}],}
where $\Gamma^{(1)}\equiv\Gamma^{12}$ and
$\Gamma^{(2)}\equiv\Gamma^{34}$. In terms of the $N=4$
supersymmetry parameters $\epsilon$, transforming in the 4 spinor
of $SO(5)_c$, the Dirac matrices $({i\over2}\Gamma^{(1)},
{i\over2}\Gamma^{(2)})$ have eigenvalues $({1\over2},{1\over2})$,
$({1\over2},-{1\over2})$, $(-{1\over2},{1\over2})$ and
$(-{1\over2},-{1\over2})$, which are simply the weights of the
spinor representation.  Using this Cartan basis, it is thus clear
that the $AdS$ black holes with 1 or 2 active charges preserve
${1\over2}$ or ${1\over4}$ of the supersymmetries respectively.

The Killing spinors may be constructed explicitly through examination of the
gravitino variation, \gvar.  Using the identity $H_\alpha^{-1}-1={1\over4}
\partial_r\log H_\alpha$ and the relation $g=2m$, we find (for the general
two charge solution)
\eqn\sgsusy{\eqalign{
\delta\hat\psi_0&=\Bigl[\partial_0-{m\over8}(\eta_1\Gamma^{(1)}
+\eta_2\Gamma^{(2)})
-{1\over2}{\cal H}^{-1/2}f\gamma_{\overline{0r}}
(\partial_r\log H_1 P^{(1)}_{\eta_1}+ \partial_r\log H_2 P^{(2)}_{\eta_2})\cr
&\qquad
+{m\over4}f^{1/2}\gamma_{\overline0} (1+r\partial_r\log H_1)P^{(1)}_{\eta_1}
+{m\over4}f^{1/2}\gamma_{\overline0} (1+r\partial_r\log H_2)P^{(2)}_{\eta_2}
\Bigr]\epsilon,\cr
\delta\hat\psi_r&=\Bigl[\partial_r+{1\over5}\partial_r\log{\cal H}
+{m\over4}{\cal H}^{1/2}f^{-1/2}(1+{r\over2}\partial_r\log{\cal H})
\gamma_{\overline r}\cr
&\qquad\qquad
-{1\over2}(\partial_r\log H_1 P^{(1)}_{\eta_1}
+\partial_r\log H_2 P^{(2)}_{\eta_2})\Bigr]\epsilon,\cr
\delta\hat\psi_{\theta_i}&=\Bigl[\partial_{\theta_i}+{1\over4}
e_{\theta_i}\gamma_{\overline{0\theta_ir}}
(\eta_1\Gamma^{(1)}+\eta_2\Gamma^{(2)})
+{1\over2}e^{\theta_j}\partial_{\theta_j}e_{\theta_i}
\gamma_{\overline{\theta_i\theta_j}}+{1\over2}f^{1/2}e_{\theta_i}
\gamma_{\overline{\theta_ir}}(P^{(1)}_{\eta_1}+P^{(2)}_{\eta_2})
\Bigr]\epsilon.
}}
Here we have assumed a standard parameterization of the unit 5-sphere, with
{\it e.g.\ }vielbeins $e_{\theta_i}=\prod_{n=1}^{i-1}\sin\theta_n$.  Solving
such Killing spinor equations is by now standard \refs{\romans\bcs}.  For
the ${1\over4}$-supersymmetric two-charge black hole we define $\hat\Gamma
={1\over2}(\eta_1\Gamma^{(1)}+\eta_2\Gamma^{(2)})$ so that the Killing
spinors may be written as
\eqn\kspin{\epsilon=e^{{mt\over2}\hat\Gamma}
e^{-{\theta_1\over2} \gamma_{\overline{0\theta_1r}}\hat\Gamma}
\left(\prod_{i=2}^5
e^{-{\theta_i\over2}\gamma_{\overline{\theta_i\theta_{i-1}}}}
\right)
\left[\sqrt{f^{1/2}+1}-\sqrt{f^{1/2}-1}\gamma_{\overline r}\right]
{\cal P}_1{\cal P}_2\epsilon_0,}
where the projections ${\cal P}_\alpha$ are individually
${1\over2}$-supersymmetry projections
\eqn\cpproj{{\cal P}_\alpha={1\over2}
(1+\eta_\alpha\gamma_{\overline0}\Gamma^{(\alpha)}).}
Note that the form of the angular components of
$\delta\hat\psi_{\theta_i}$ ensures that the angular part of
\kspin\ corresponds to modified Killing spinors on the sphere \lpr.

The above form of the Killing spinor equations, obtained for the general
two-charge solution, is presented in a symmetric manner; it is no longer
fully appropriate for the case of a ${1\over2}$-supersymmetric solution,
corresponding to a single active charge.  Nevertheless, assuming that the
single charge is carried by $F_{\mu\nu}^{(1)}$ (so that $H_2=1$), it is
straightforward to remove terms dependent on $P^{(2)}$ and $\Gamma^{(2)}$
from \kspin.  The resulting Killing spinors are given by \kspin\ after
dropping ${\cal P}_2$ and with the replacement
$\hat\Gamma\to\eta_1\Gamma^{(1)}$.

\newsec{Membranes in $AdS_7$}
Proceeding from the analysis of the black holes in $AdS_7$, we now wish
to investigate the possibility of constructing extended objects in an $AdS$
background.  In particular, we now consider membranes which are
electrically charged with respect to $C_3$. To the best of our knowledge,
all solutions discussed so far in the $AdS$ literature have the
antisymmetric tensor fields which are in the
fundamental representation of the $R$-symmetry group switched
off.  Turning on such fields poses several difficulties which apparently
have not yet been fully addressed.  These include understanding the nature
of the odd-dimensional self-duality equation \ceom\ as well as the
construction of Killing spinors appropriate to the brane in an $AdS$
background.

As as first step in constructing a membrane solution with an active $C_3$
field, we set the gauge fields to zero (they will be reexamined
subsequently).  In this case it is natural to expect the $SO(4)$ symmetry to
be restored, so that the $R$-symmetry breaking is given simply by
$SO(5)_g\to SO(4)$ without the further breaking to $U(1)^2$.  This may be
accomplished by setting $\lambda_1 = \lambda_2 = \lambda$, resulting
in the further truncation of the potential ${\cal V}$ to
\eqn\tpot{{\cal V}=-8e^{4\lambda}-8e^{-6\lambda}+e^{-16\lambda}.}
While the three-form equation of motion was given in \ceom, we note that it
may be rewritten in second-order form as
\eqn\ctwo{\nabla^\alpha(e^{8\lambda}G_{\alpha\beta\gamma\delta})
=-{m\over24}\epsilon_{\alpha\beta\gamma\delta}{}^{\mu\nu\lambda}
G^\alpha{}_{\mu\nu\lambda},}
where $G_{\alpha\beta\gamma\delta}=4\partial_{[\alpha}C_{\beta\gamma\delta]}$.
Note however that the usual antisymmetric tensor gauge invariance is lacking
in \ctwo, as the bare potential must still satisfy the first order
equation \ceom.  In this form, the odd-dimensional self-duality condition
indicates that $G_4$ provides its own source.  In fact, written as
$d(e^{8\lambda}*G)=mG$, we see that the ``magnetic'' components of $G_4$
serve as its electric source.  As a result, the usual distinction between
electric and magnetic 2- and 1-branes respectively is not so clear in this
instance.

The membrane ansatz implies a split of the coordinates into three
longitudinal and four transverse dimensions.  Writing the transverse space
as $R^+\times S^3$ suggests a natural choice of the line element to be
of the form
\eqn\choi{ds^2=e^{2A}[-dt^2+d\vec x^2] + e^{2B} dr^2 + e^{2C} r^2d\Omega_3^2,}
with $\vec x$ denoting the spatial longitudinal coordinates $\{x_1,x_2\}$.
The resulting split suggests an ansatz for the three-form in which its
non-vanishing components are $C_{012}$ (electric) and
$C_{\theta_1\theta_2\theta_3}$ (magnetic); these two components are further
related by the self-dual equation of motion \ceom.

At this stage one immediately runs into a difficulty in that the metric
\choi\ does not easily admit a vacuum $AdS_7$ background.  This issue
arises from the separation of the three coordinates $\{t,x_1,x_2\}$ to form
the longitudinal directions of the membrane; this split destroys the overt
symmetry among the six spatial dimensions (and $AdS$ is maximally symmetric).
Physically, this corresponds to the notion that {\it static} extended
objects do not appear to exist in anti-de Sitter geometries.  Similar
issues have already been discussed for $AdS$ black holes, where no examples
of multi-center black hole solutions in an $AdS$ background have yet been
obtained.

While it remains an open problem to reconcile the symmetry of the $AdS$
background with that of an extended object, we may nevertheless study the
characteristics of the membrane solution in an asymptotic regime.  In this
manner we are able to provide a closer investigation of the first order
self-duality equation \ceom.  Far from the membrane we expect the geometry
of spacetime to be asymptotically $AdS_7$.  Examining the minimum of the
potential \tpot\ indicates the presence of a cosmological constant
corresponding to $R_{\mu\nu\lambda\sigma}=-{m^2\over4}(g_{\mu\lambda}
g_{\nu\sigma}-g_{\mu\sigma}g_{\nu\lambda})$.  Furthermore, as $r\to\infty$,
the $S^3$ asymptotically ``flattens out'' \wtpt, so that the line element
\choi\ may take on a horospherical form
\eqn\horo{ds^2= {m^2r^2\over4}[-dt^2+d\vec x^2] + {4\over m^2r^2}dr^2
+ {m^2r^2\over4}d\Omega_{3,0}^2.}
Here $d\Omega_{3,0}^2=d\theta_1^2+d\theta_2^2+d\theta_3^2$ denotes the flat
metric on $T^3$.  This modification to the consideration of non-spherical
horizons is a possibility of non-asymptotically-flat spacetimes, and has
been used in the investigation of $AdS$ black holes
\refs{\cdk,\dk,\bcvs,\cg,\tena}.

In the background of \horo, the self-dual equation of motion now takes on
the ``symmetric'' form
\eqn\sdho{\eqalign{
C_{012}&=-{1\over2}e^{8\lambda}r\partial_rC_{\theta_1\theta_2\theta_3},\cr
C_{\theta_1\theta_2\theta_3}&=-{1\over2}e^{8\lambda}r\partial_rC_{012},}}
and hence yields
\eqn\twi{C_{012}={1\over4}e^{8\lambda}r\partial_r(e^{8\lambda}
r\partial_rC_{012}),}
when iterated twice.  Assuming $\phi\sim 1/r^p$ for $p>1$, the above may be
solved asymptotically to yield
\eqn\asyc{C_{012}\sim{Q\over r^2},\qquad
C_{\theta_1\theta_2\theta_3}\sim{Q\over r^2},}
indicating that both components are of equal importance.  In fact, this
results in a cancellation between ``electric'' and ``magnetic'' terms so
that $C_{\mu\nu\lambda}^2\sim (Q^2/ m^6) O(r^{-(10+p)})$.  This has direct
correspondence to the vanishing of $F_{2k+1}^2$ for self-dual $(2k+1)$-forms
in $4k+2$ dimensions.  As a result, at least to this order,
$C_{\mu\nu\lambda}^2$ drops out of the scalar equation of motion, \seom, so
that $\lambda$ may be set to zero, indicating the consistency of such
asymptotics.  Examination of the Einstein equation, \eins, finally reveals
the distinction between the longitudinal and transverse directions
resulting from the relative minus sign between $C_{012}^2$ and
$C_{\theta_1\theta_2\theta_3}^2$.  Presumably this gives rise to the
appropriate powers of ``harmonic functions'' that naturally appear in
$p$-brane ans\"atze when the back-reaction of the test-membrane \asyc\ is
taken into account.

In the absence of a test-membrane, the $AdS$ line element \horo\
admits Killing spinors \refs{\lpt,\lpr} $\epsilon_\pm$ satisfying
both projections $P_\pm={1\over2}(1\pm\gamma_{\overline r})$.  As
an asymptotic solution, we note that the Killing spinor equations
arising from \gvar\ may no longer be satisfied identically, but
only up to terms of $O(r^{-5})$ compared to the $AdS$ background.
On the other hand, for $\delta\lambda$, we obtain {\it at leading
order}
\eqn\lol{\eqalign{\delta\lambda&\sim{2\sqrt{3}\over m^2r^3}
(C_{012}\gamma^{\overline{012}}+C_{\theta_1\theta_2\theta_3}
\gamma^{\overline{\theta_1\theta_2\theta_3}})\Gamma^5\epsilon\cr
&\sim{2\sqrt{3}\over m^2r^3}\gamma^{\overline{012}}\Gamma^5
(C_{012} + C_{\theta_1\theta_2\theta_3}\gamma_{\overline r})\epsilon\cr
&\sim{4\sqrt{3}Q\over m^2 r^5}\gamma^{\overline{012}}\Gamma^5
P_+\epsilon.}}
Therefore this indicates as expected that the membrane preserves only
half of the
original supersymmetries---namely those corresponding to $P_+\epsilon=0$.

\newsec{Discussion}

The seven-dimensional $N=4$ gauged supergravity \refs{\ppvn,\ppvnw}\
admits a rich structure, both in terms of symmetries and of solutions.  In
fact a curious feature of this theory is that, unlike gauged maximal
supergravities in $D=4$ and $D=5$, here the potential admits not only a
$SO(5)$ stationary point, but also a $SO(4)$ saddle point.  It is curious
that this latter point also appears to correspond naturally to the
symmetry breaking induced by a single $C_3$-charged membrane.

While we have provided the asymptotic behavior of a membrane in an $AdS$
background, the full solution is still lacking.  The difficulty in the
construction of such a solution appears to lie in the presence of extra
longitudinal directions: viewed from the worldvolume, these ought to be
``flat'', while viewed from $AdS$ spacetime these ought to be ``curved''.
This contradiction is similar in spirit to one that arises when searching
for multi-center black hole solutions in $AdS$.  In that case there is
difficulty in resolving the BPS no-force condition with the curvature of
space.

One possible approach to investigating extended $p$-brane solutions involves
the use of non-static metric ans\"atze.  Alternatively, one may involve
dependence of the metric on more than a single coordinate.  For example one
possibility is a metric in the modified horospherical form
\eqn\metric{ds^2 = e^{2A+2z} (-dt^2 + dx^2) + e^{2B}dz^2 + e^{2C+2z}
dy^i dy^j\delta_{ij},}
where $A$, $B$ and $C$ are functions of $r=|\vec y|$.  While the rationale
behind this choice is to separate out the $AdS$ coordinate $z$ from the
brane transverse coordinate $r$, this has in practice not yet yielded any
further solutions.

Our final comment on the membrane solution regards the gauge
fields that so far were set to zero. To turn these on, it appears we
do not want to work in the $U(1)^2$ truncation presented in section 2.
Instead it is important to keep the full $SO(4)\subset SO(5)_g$ symmetry
(always taking $\lambda_1=\lambda_2=\lambda$) and to retain the non-abelian
$SO(4)$ gauge fields $A_\mu^{IJ}$ where $I,J=1,2,3,4$.  In this case the
resulting lagrangian and equations of motion are easily generalized from
those given in \trunc--\eins.  The supersymmetry variations are similarly
obtained.  In particular, the spin-${1\over2}$ variations, \dvar, collapse
to give the single variation
\eqn\navar{\delta\lambda=\Bigl[{m\over4}(e^{2\lambda}
-e^{-8\lambda})-{5\over4}\gamma^\mu\partial_\mu\lambda
-{1\over8}\gamma^{\mu\nu}e^{-2\lambda}F_{\mu\nu}^{IJ}
\Gamma^{IJ} +{m\over8\sqrt{3}}\gamma^{\mu\nu\lambda}
e^{-4\lambda}C_{\mu\nu\lambda}\Gamma^5\Bigr]\epsilon.}
Interestingly enough, for a gauge field that is (anti-)self-dual in both
the transverse space SO(4) and the internal SO(4) directions,
\eqn\inst{
F_{ij}^{IJ}=\pm{1\over4}\epsilon_{ij}{}^{kl}\epsilon^{IJKL}F_{kl}^{KL},
}
we find
\eqn\instvar{
\gamma^{ij}\Gamma^{IJ}F_{ij}^{IJ}
={1\over2}F_{ij}^{IJ}\gamma^{ij}\Gamma^{IJ}
(1\pm\gamma^{\overline{3456}} \Gamma^{1234}),}
where $i,j=3,4,5,6$ denote transverse space directions.  By dualizing
the above combination of Dirac matrices, this combination takes on the form
of a standard electric 2-brane projection, $\widetilde P_\pm$, where
\eqn\iproj{\widetilde P_\pm ={1\over2}(1\pm\gamma^{\overline{012}}\Gamma^5).}
Note, however, that while this is the form of a ${1\over2}$-supersymmetry
projection that would be used to construct a supersymmetric membrane
solution in the {\it ungauged} supergravity, it is nevertheless distinct
from the $P_\pm$ obtained in the asymptotic limit of the previous section.

Based on analogy with the ungauged theory, it appears that a complete
membrane solution would have to incorporate a supersymmetry projection
more of the form of $\widetilde P_\pm$ than of the form $P_\pm$.  Thus we
conjecture that it is in general possible to turn on the gauge fields by
taking an instanton configuration \inst\ transverse to the membrane
(both in spacetime and in the ``$R$-directions") without breaking any extra
supersymmetry. It is conceivable that, in analogy with the $AdS_5$
case \ferp, there are other breakings of the $R$-symmetry
corresponding to BPS states preserving less than half of the
supersymmetry. It would certainly be interesting to see what gauge
configurations these would correspond to.

Finally, we point out that the membrane solution in $AdS_7$ has the
interpretation of the near $M5$-brane limit of the $M5$ intersect
$M2$ system in eleven dimensions.  While the complete $M5\perp M2$
system preserves only ${1\over4}$ of the supersymmetries, in the
near-horizon limit of the $M5$-brane this is restored to
${1\over2}$ of the supersymmetries.  We have thought about taking
advantage of this decoupling limit in constructing the membrane
solution from an eleven-dimensional starting point.  However fully
localized intersecting brane solutions still appear quite obscure
\refs{\intbr,\jpg,\ity,\youm}.
Even in the decoupling limit of one of the branes, it appears
that some delocalization in one of the relative transverse
directions remains necessary (as a technical simplification).  Such
considerations are presumably related to the issues that must be
addressed in performing a direct construction of the membrane
solution in $AdS_7$.

\bigskip
\centerline{\bf Acknowledgments}\nobreak
\bigskip

The work of JTL is supported in part by grant DE-FG02-91ER40651-TASKB; the
work of RM is supported by DOE grant DE-FG02-92ER40704.

\listrefs
\bye